\documentclass[preprintnumbers,amsmath,amssymb, twocolumn]{revtex4}
\usepackage{graphicx}
\usepackage{dcolumn}
\usepackage{bm}
\usepackage{braket}
 \usepackage{mathrsfs}
 \usepackage{amsmath}

\begin{document}

\title{Coherent Population Trapping with a controlled dissipation: \\  applications in optical metrology}

\author{L. Nicolas$^{1}$}
\author{T. Delord$^{1}$}
\author{P. Jamonneau$^{2}$}
\author{R. Coto$^{4}$}
\author{J. Maze$^{3}$}
\author{V. Jacques$^{2}$}
\author{G. H\'etet$^{1}$} 

\affiliation{$^1$Laboratoire Pierre Aigrain, Ecole normale sup\'erieure, PSL Research University, CNRS, Universit\'e Pierre et Marie Curie, Sorbonne Universit\'es, Universit\'e Paris Diderot, Sorbonne Paris-Cit\'e, 24 rue Lhomond, 75231 Paris Cedex 05, France. \\
$^2$ Laboratoire Charles Coulomb, Universit\'e de Montpellier and CNRS, 34095 Montpellier, France. \\
$^3$ Departamento de Fisica, Pontificia Universidad Catolica de Chile, Casilla 306, Santiago, Chile. \\
$^4$ Universidad Mayor, Avda. Alonso de C\'ordova 5495, Las Condes, Santiago, Chile.
}
\begin{abstract}
We analyze the properties of a pulsed Coherent Population Trapping protocol that uses a controlled decay from the excited state in a $\Lambda$-level scheme. We study this problem analytically and numerically and find regimes 
where narrow transmission, absorption, or fluorescence spectral lines occur. 
We then look for optimal frequency measurements using these spectral features by computing the Allan deviation in the presence of ground state decoherence and show that the protocol is on a par with Ramsey-CPT. We discuss possible implementations with ensembles of alkali atoms and single ions and demonstrate that typical pulsed-CPT experiments that are realized on femto-second time-scales can be implemented on micro-seconds time-scales using this scheme.
\end{abstract}

\maketitle

Since its observation \cite{Gray78}, Coherent-Population-Trapping (CPT) and its counterpart Electromagnetically-Induced-Transparency (EIT) \cite{Fleischhauer}, have enabled a wide range of 
experimental achievements. Using a three-level $\Lambda$-scheme and exploiting a quantum interference effect in the excited state, coherent transfer of population (STIRAP) between vibrational states \cite{Ni231}, efficient cooling of atoms \cite{Aspect}, precise atomic clocks \cite{Vanier2005, Hafiz}, or light storage \cite{Phillips, liu} have been realized. Ramsey-CPT schemes have also been shown to improve the sensitivity of frequency measurements by removing power broadening issues \cite{Hafiz}.
Further, pulsed-CPT schemes have been investigated theoretically \cite{Kocharovskaya, Moreno11, Soares, Soares3, Soares2, Ilinova} and experimentally using femto-second lasers \cite{Sautenkov, Arissian, Campbell} with implications for multimode quantum memories \cite{Simon}.

Recently, a novel pulsed-CPT scheme was realized by engineering a $\Lambda$-system in the microwave domain and exploiting the hyperfine interaction between the electron spin of a nitrogen-vacancy (NV) defect in diamond and a nearby $^{13}$C nuclear spin  \cite{Jamonneau}.
The originality of the experiment is that relaxation was externally controlled through optical pumping by a far detuned laser that couples the excited state to the ground state in the $\Lambda$-system via a metastable state in the NV center level structure. 
In general such a scheme with controlled relaxation is useful for atomic systems where the excited state population lifetime is too long compared to the decoherence mechanisms.
Further, compared to schemes where spontaneous emission takes place during the excitation, this method can be used to measure the excited state population while the dark state is being prepared.

\begin{figure}[h!]
\centerline{\scalebox{0.2}{\includegraphics{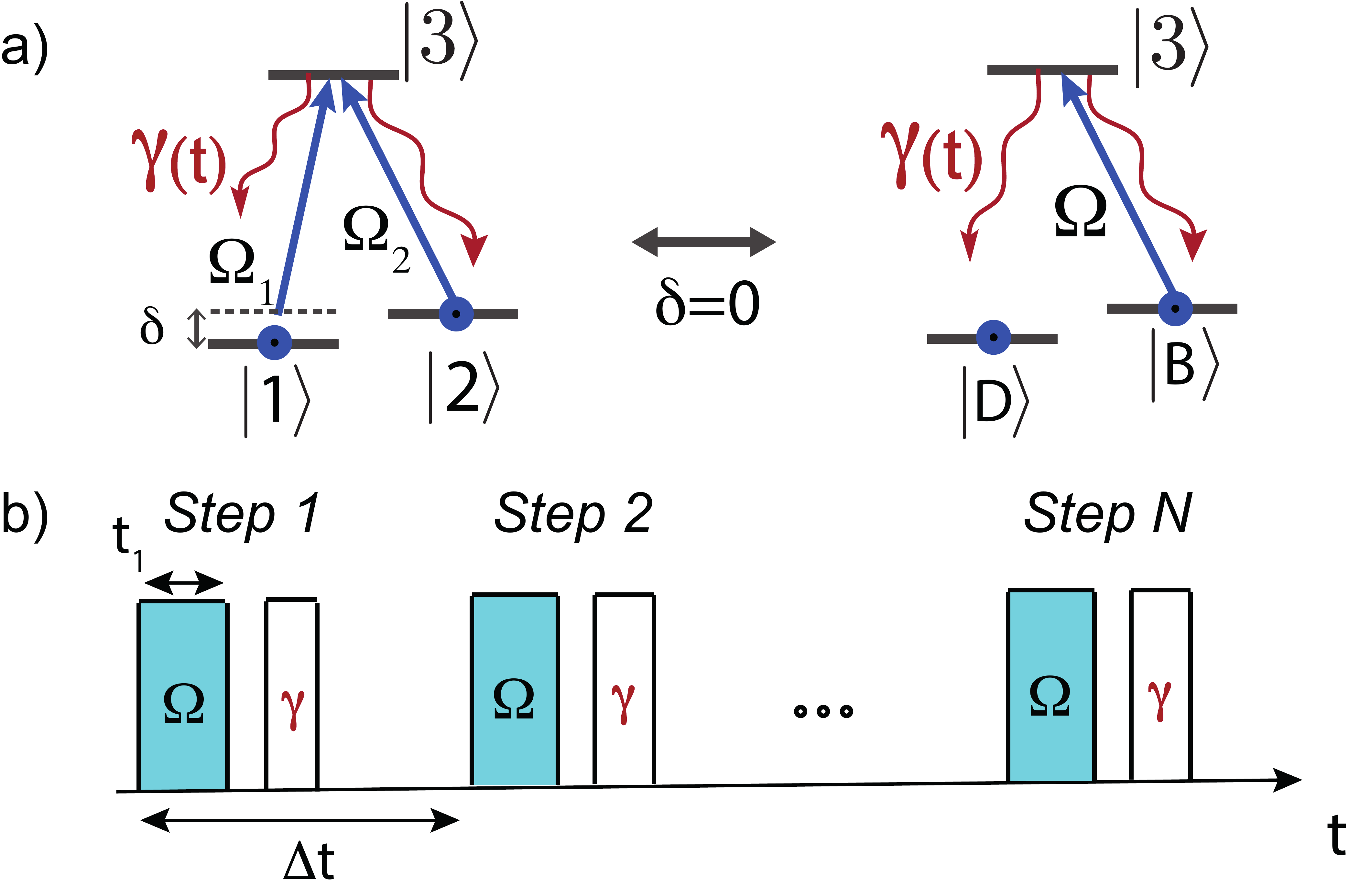}}}
\caption{a) Three level $\Lambda$-system in the initial and dressed state bases with a time dependent decay rate. b) Sequence used for pulsed-CPT with controlled dissipation.
}
\label{Lambda}
\end{figure}

In this work, we analyze such a pulsed-CPT scheme with a controlled decay from the excited state, both analytically and numerically.  We show that interleaving sequences of unitary and fully dissipative steps gives rise to narrow dark transmission and photoluminescence spectral lines whose widths do not depend upon the spontaneous emission rate but solely upon the control pulse area and number of steps. 
Then, we discuss implications of this scheme for metrology and estimate its precision compared to Ramsey-CPT. Finally, we present possible experimental implementations using neutral alkali atoms or trapped ions, and show how pulsed-EIT experiments that are typically realized using femto-second lasers can be implemented on micro-seconds time-scales. 

\begin{figure}[h]
\centerline{\scalebox{0.28}{\includegraphics{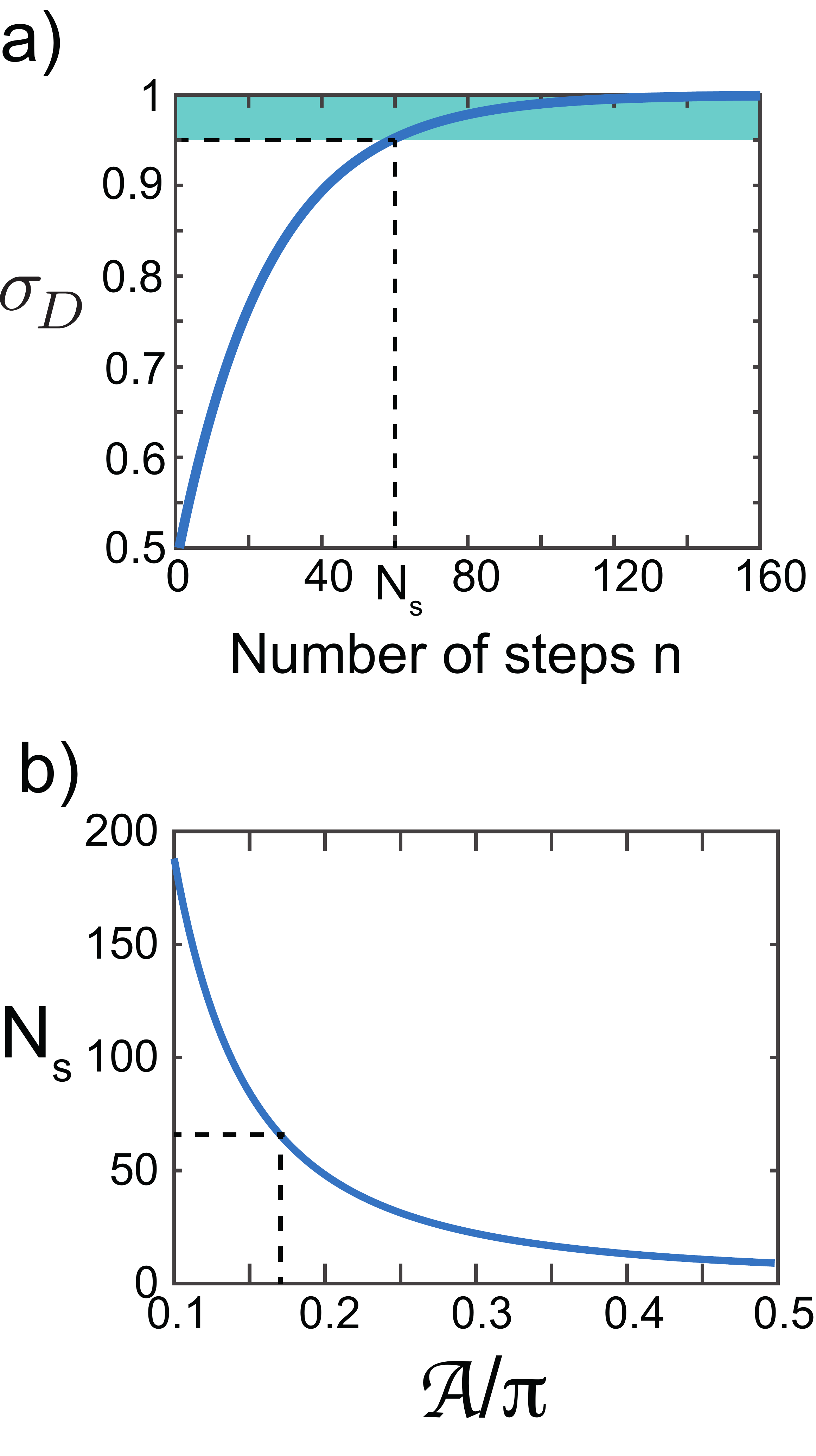}}}
\caption{
a) Population in the state $\ket{D}$ as a function of the number of steps starting from a thermal population $\sigma_{D}^0=0.5$ and using a pulse area $\mathscr{A}=0.18\pi$. The grey region denotes the threshold for reaching more than 95\% population in the dark state. 
b) Number of steps needed to reach the final dark state population $\sigma_{Df}=0.95$ as a function of the pulse area.
}
\label{fig4}
\end{figure}

We study the response of a three-level $\Lambda$-system excited by two near-resonant pulsed fields with Rabi frequencies $\Omega_1$ and $\Omega_2$ in the presence of a controlled relaxation rate $\gamma(t)$, as sketched in Fig. 1-a). Throughout the manuscript the field $2$ will be resonant and we look for the change in transmission of the field $1$ or of the excited state population as the frequency detuning of the field $1$, $\delta=\omega_1-\omega_{13}$ is varied. Here $\omega_1$ denotes the frequency of the field 1 and $\omega_{13}$ the resonance frequency of the transition $\ket{1}-\ket{3}$.
The sequence is composed of a periodically pulsed simultaneous excitation of the two transitions interleaved by controlled relaxation, as depicted Fig. 1-b). One sequence is composed of $N$ pulses separated by a time interval $\Delta t$. $t_1$ is the length of the excitation pulses. 

\section{Pulsed-CPT under resonant excitation}

At two-photon resonance ($\delta=0$), one can write the $\Lambda$-system using a basis consisting of a dark state $\ket{D}=\frac{1}{\Omega}(\Omega_2\ket{1}-\Omega_1\ket{2})$ which is an eigenstate of the coupled $\Lambda$-system \cite{Fleischhauer}, a bright state $\ket{B}=\frac{1}{\Omega}(\Omega_1\ket{1}+\Omega_2\ket{2})$ where $\Omega=\sqrt{\Omega_1^2+\Omega_2^2}$, and the excited state, as sketched in Figure 1-a).  The dark state $\ket{D}$ is not coupled to the excited state whereas the bright state $\ket{B}$ couples to the excited state with a Rabi pulsation $\Omega$, and enables optical pumping to $\ket{D}$. 

Neglecting decoherence ($\gamma_0=0$), the $n^{\rm th}$ coherent excitation pulse transfers a population $(\sin\frac{\mathscr{A}}{2})^2\sigma_{B}^{n-1}$ from the bright state to the excited state, where $\mathscr{A}=\Omega t_1$ is the pulse area and 
$\sigma_{B}^{n-1}$ is the population in the bright state just before the $n^{\rm th}$ pulse. 
Then, the induced decay transfers half of this quantity from the excited state population to the dark state and the other half to the bright state. The population in the dark state at the $n^{\rm th}$ steps is described by the equation 
$\sigma_D^{n}=\frac{1}{2}(\sin\frac{\mathscr{A}}{2})^2(1-\sigma_D^{n-1})+\sigma_D^{n-1}$
which can be solved to give $\sigma_D^{n}=1-\left(1-(\sin\frac{\mathscr{A}}{2})^2/2\right)^{n-1}\left(1-\sigma_D^0\right)$.

The dark state population is plotted in Fig. 2-a) as a function of $n$ starting from the thermal population in the bright state $\sigma_{B}^0=0.5$ and using a pulse area $\mathscr{A}=0.18\pi$. Here $N_s=60$ steps are needed to reach a final dark state occupation above $0.95$. 
We then plot the dependence of $N_s$ on the pulse area in Fig. 2-b). 
We find that, as expected, using shorter pulses (smaller $\mathscr{A}$) means that more pulses are required to reach the same dark state population.

\section{Pulsed-transmission with controlled decay}

We now proceed with the study of the frequency response of such a pulsed-CPT scheme. 
Experimentally, the imaginary part of the dipole amplitude $\sigma_{13}=|1\rangle \langle 3|$ is relevant for transmission experiments because it relates to the imaginary part of the linear susceptibility for the field $1$. Here we assume that the field $1$ is much weaker than field $2$, $\Omega_1\ll\Omega_2$, which greatly simplifies the analysis. Since it is much weaker, the field $1$ will hereafter be called the probe and the field $2$ the control.
In order for the probe transmission to be modified, a macroscopic number of atoms, an optical cavity, or a quasi-1D geometry must be used. 
With many atoms, all dipole operators $\sigma_{ij}$ would be locally averaged \cite{Gorshkov} \footnote{If the medium is optically thick, coupled Maxwell-Bloch equations must be used, as in \cite{Soares}. Retardation effects would take place, but the main conclusions of the paper remain unchanged.}. In the case of a single atom in a cavity or with a quasi-1D geometry, the dipole coupling strength would be rescaled by the cavity or waveguide spatial modes. 

To study this problem, we solve the time-dependent Bloch equations. 
In the Heisenberg picture, the evolution of the mean value of the Pauli operators is ruled by the 6 optical Bloch equations that are written in the appendix (\ref{eqBloch}). 
Assuming that $\Omega_1\ll\Omega_2$, the population remains in the state $\ket{1}$ and the other states are not significantly populated. 
Under these hypotheses we find
\begin{equation}
\label{eqBlochsimp}
	    \begin{array}{l} 
 	       \dot{\sigma}_{13}=(-\gamma(t)+i\delta)\sigma_{13}+i\frac{\Omega_1}{2}+i\frac{\Omega_2}{2}\sigma_{12} \\
 	       \dot{\sigma}_{12}=(i\delta+\gamma_0)\sigma_{12}+i\frac{\Omega_2}{2}\sigma_{13}, \\ 
 	   \end{array}
\end{equation}
where $\delta$ is the detuning of the field $\Omega_1$, $\sigma_{13}=|1\rangle\langle3|$ and $\sigma_{12}=|1\rangle\langle2|$ which can be treated as c-numbers in this approximate regime. $\gamma_0$ is the dephasing rate of the ground state coherence. We can solve analytically this system of equations under the assumption of full relaxation, which implies that $\sigma_{13}=0$ at the end of each excitation pulse. 

\subsection{Multiple interferences}

We study ${\rm Im}(\sigma_{13})$ as a function of the two-photon detuning $\delta$ in the absence of ground state dephasing $\gamma_0$. Let us first note that the evolution of $\sigma_{13}$ during the $n^{th}$ excitation pulse is directly related to the value of $\sigma_{12}$ at the beginning of the pulse $\sigma_{12}^n=\sigma_{12}\left((n-1)\Delta t\right)$. Solving Eqs. (\ref{eqBlochsimp}) yields a simple formula for $\sigma_{12}^n$. We get, for $n>1$:

\begin{equation}
\label{s12n}
	   \sigma_{12}^n(\delta)=f(\delta)\sum_{l=0}^{n-2} e^{i\delta l\Delta t}\cos^l\left(\frac{\mathscr{A}}{2}\right),
\end{equation}

with \[\begin{array}{l}
 	      f(\delta)=\frac{1}{2}\frac{\Omega_1 e^{i\delta\Delta t} }{\delta^2-(\frac{\Omega_2}{2})^2}\left(\frac{\Omega_2}{2}\left(\cos(\delta t_1)-\cos\left(\frac{\mathscr{A}}{2}t_1\right)\right)\right.   \\
 	      \mbox{  \qquad \qquad \qquad} \left.+i\left(\delta \sin\left(\frac{\mathscr{A}}{2}t_1\right)-\frac{\Omega_2}{2} \sin(\delta t_1)\right)\right). \\

  	   \end{array}
\]

We observe that $\sigma_{12}^n$ is proportional to a geometric sum, analogous to the transmission of light in a Fabry-P\'erot cavity. We thus expect to observe a periodic spectrum with interference peaks separated by the free spectral range defined as $\Delta\delta_{FSR}=\frac{2\pi}{\Delta t}$.
The width of each peak should thus decrease with the number of steps until a steady state is reached.
When $N$ is high enough, a simple formula including $\gamma_0$ can be found, and the second term in the product appearing in Eq. 2 will show peaks with a width related to a finesse given by \[\mathscr{F}=\frac{\pi\sqrt{e^{-\gamma_0 \Delta t}\cos\left(\frac{\mathscr{A}}{2}\right)}}{1-e^{-\gamma_0 \Delta t}\cos\left(\frac{\mathscr{A}}{2}\right)}.\] This formula is valid for $\cos\frac{\mathscr{A}}{2}>1/2$, {\it i.e}. for $\mathscr{A}<2\pi/3$. The width of each peak is then $\mathscr{F}/\Delta t$.
We note that $\mathscr{F}$ increases when $\mathscr{A}$ decreases, as shown Fig. 3-a). 


As manifest in Eq. (2), there are in fact two distinct multiple-interference regimes. A regime for which $\cos\left(\frac{\mathscr{A}}{2}\right)\approx 1$, where constructive interferences occur when $\delta \Delta t=2m\pi$ ($m \in \mathbb{Z}$) and a regime in which $\cos\left(\frac{\mathscr{A}}{2}\right)\approx -1$, where constructive interferences take place at $\delta \Delta t=(2m+1)\pi$. 
We will study the physics underlying these two situations in the next sections by evaluating $\sigma_{13}$ when the area is smaller then $2\pi$ and when the area is close to $2\pi$.

\subsection{Pulsed-EIT regime}

\begin{figure}[h]
\centerline{\scalebox{0.30}{\includegraphics{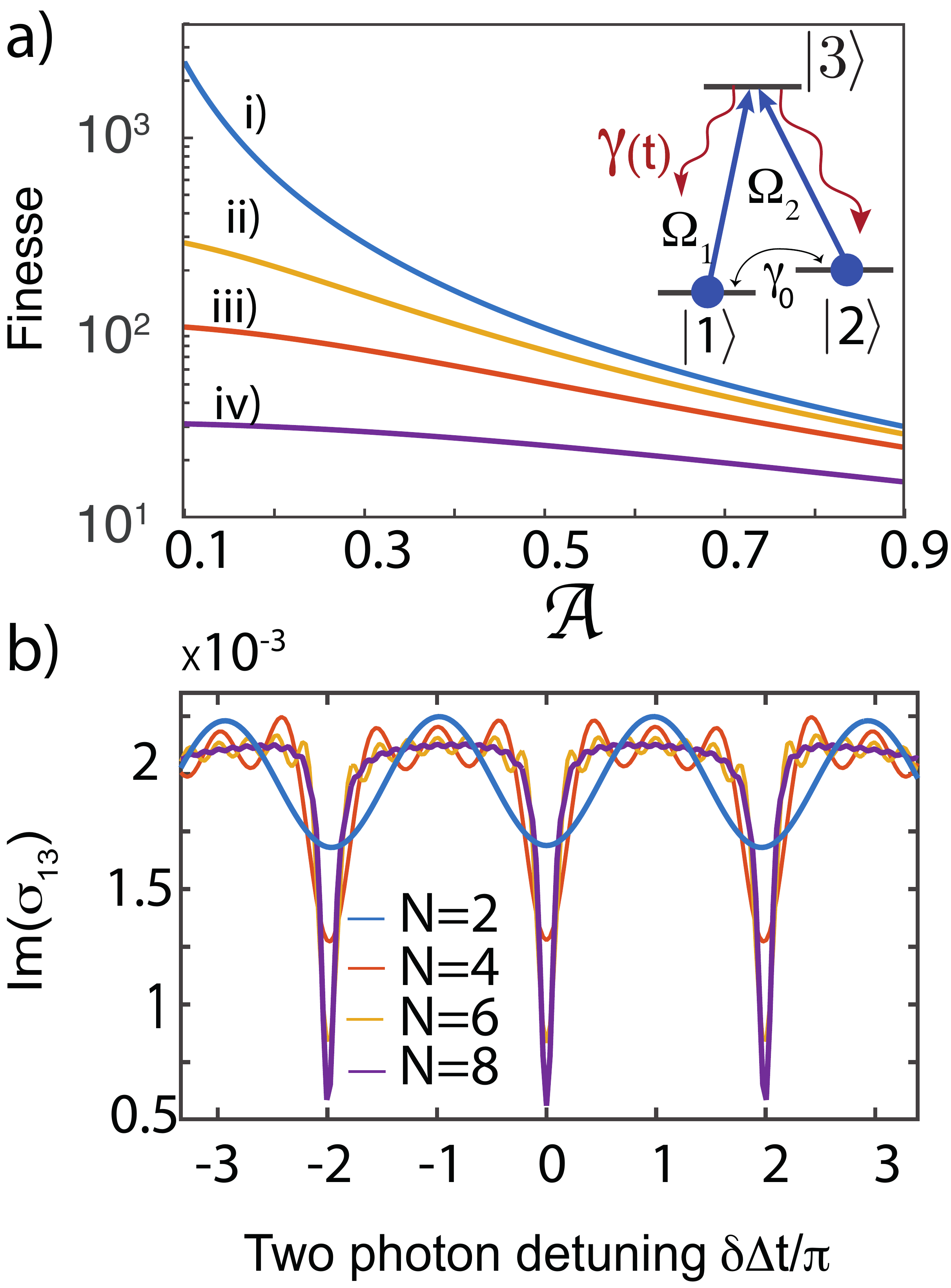}}}
\caption{
a) Finesse of the EIT transmission peaks as a function of the pulse area $\mathscr{A}$ for $\gamma_0 \Delta t=$ 0, 0.01, 0.03, 0.06, for trace i) to iv) respectively.
b) Transmission spectra for a different number of preparation pulses $N$. Parameters: $\Omega_1/2\pi=31.8$ kHz, $\Omega_2/2\pi=6.37$ MHz, $\Delta t=1\ \mu s$ and $\mathscr{A}=2\pi/5$. }
\label{fig4}
\end{figure}

We first study the transmission spectra using the analytical solution of the Bloch equations A-\ref{eqBlochsimp} derived in the appendix, in the limit where the control pulse area is below $2\pi$.
Transmission spectra measured at the end of the last pulse of the sequence are plotted as a function of the two-photon detuning for a different number of steps $N$ in Fig. 3-b).
The parameters are $\Omega_1/2\pi=31.8$ kHz, $\Omega_2/2\pi=6.37$ MHz, $\Delta t=1 \ \mu s$ and $\mathscr{A}=2\pi/5$. 
As anticipated, a periodic pattern appears in the transmission spectrum. Close to two-photon resonance, we observe a spectrum that is similar to EIT and we clearly see a step-by-step narrowing of the spectrum.  One sees that the central peak amplitude gets closer and closer to zero after several pulses. The multiple interference in the accumulated ground state coherence also implies that the width of the peaks narrows down as the number of pulses $N$ increases, as per Eq. 2
and that other transmission peaks that are equally spaced at frequencies $\delta\Delta t=2m\pi$ where $m \in \mathbb{Z}$ appear due to this frequency comb driving as expected. 
This confirms that there is a sequential pumping into the dark state $\ket{D}=\frac{1}{\Omega}(\Omega_2\ket{1}-\Omega_1\ket{2})$.  Indeed, adding detunings to both fields only changes the local phase in $\ket{D}$ to $\exp(\delta \Delta t)$ which remains an eigenstate if $\delta\Delta t=2m\pi$.

\subsection{Pulsed Autler-Townes regime}

Let us now investigate the regime where the pulse area is $2\pi$. 
When $\mathscr{A}=2\pi$, the population in the Bloch vector undergoes a full rotation on the Bloch sphere related to the $\ket{2}$-$\ket{3}$ transition. 
The probe transmission spectrum, measured right after the excitation pulses would thus be independent on the probe frequency.
In order to optimally measure the ground state coherence evolution in this regime, the probe transmission must thus be acquired in the middle of the coherent excitation, {\it i.e.} on the north pole of the Bloch sphere, as depicted in Fig. 4-a).

Fig. 4-b), trace i) shows the evolution of ${\rm Im} (\sigma_{13})$ as a function of the two photon detuning, in the steady state and small area limit. The parameters are the same as for Fig. 2-b).
Fig. 4-b), trace ii) shows the same spectrum for $\Omega_1/2\pi=31.8$ kHz, $\Omega_2/2\pi=6.37$ MHz, $N=15$ and $\Delta t=1 \ \mu s$, that is for a pulse area of $2\pi$. 
The difference between the two excitation regimes shown in trace i) and ii) is striking : instead of transmission peaks at detunings such that $\delta\Delta t=2m\pi$ with $m \in \mathbb{Z}$, absorption peaks occur for $\delta\Delta t=(2m+1)\pi$. 

This $\mathscr{A}=2\pi$ regime is in fact reminiscent of the Autler-Townes effect where, under continuous wave excitation (CW) and under saturation ($\Omega_2\gg \gamma$), an absorption doublet appears at $\delta=\pm \Omega_2/2$ \cite{AT}.
Here, when $\mathscr{A}=2\pi$, $\sigma_{13}=0$ at the end of each excitation pulse, so the induced decay has no effect on the coherences.
To understand why the absorption peaks occur at $\delta\Delta t=(2m+1)\pi$ then, it is instructive to define an effective Rabi frequency $\Omega_{\rm eff}$ for the whole duration $\Delta t$, so as to tend and to eventually compare to the quasi-continuous driving (CW). Defining a Rabi frequency for a duration $\Delta t$ is possible here because controlled decay has no influence when $\mathscr{A}=2\pi$.  
We define $\Omega_{\rm eff}$ as the Rabi frequency that would populate the excited state during a time $\Delta t$ with the same probability as a shorter excitation at a Rabi frequency $\Omega_2$ for a time $t_1$. Equating the pulse areas, we obtain $\Omega_{\rm eff}\Delta t=\Omega_2 t_1$.  Since $\mathscr{A}=2\pi$ here, we get $\Omega_{\rm eff}=2\pi /\Delta t$.
The two first absorption peaks appear at $\delta=\pm\pi/\Delta t$ which means $\delta=\pm\Omega_{\rm eff}/2$. 
This analysis shows that the doublets appear at the same frequencies as the CW Autler-Townes doublets. This also shows that the transition from EIT to Autler-Townes, also investigated in \cite{giner, Anisimov}, can be observed using a $\Lambda$-scheme with controlled dissipation. 

\begin{figure}[h!]
\centerline{\scalebox{0.16}{\includegraphics{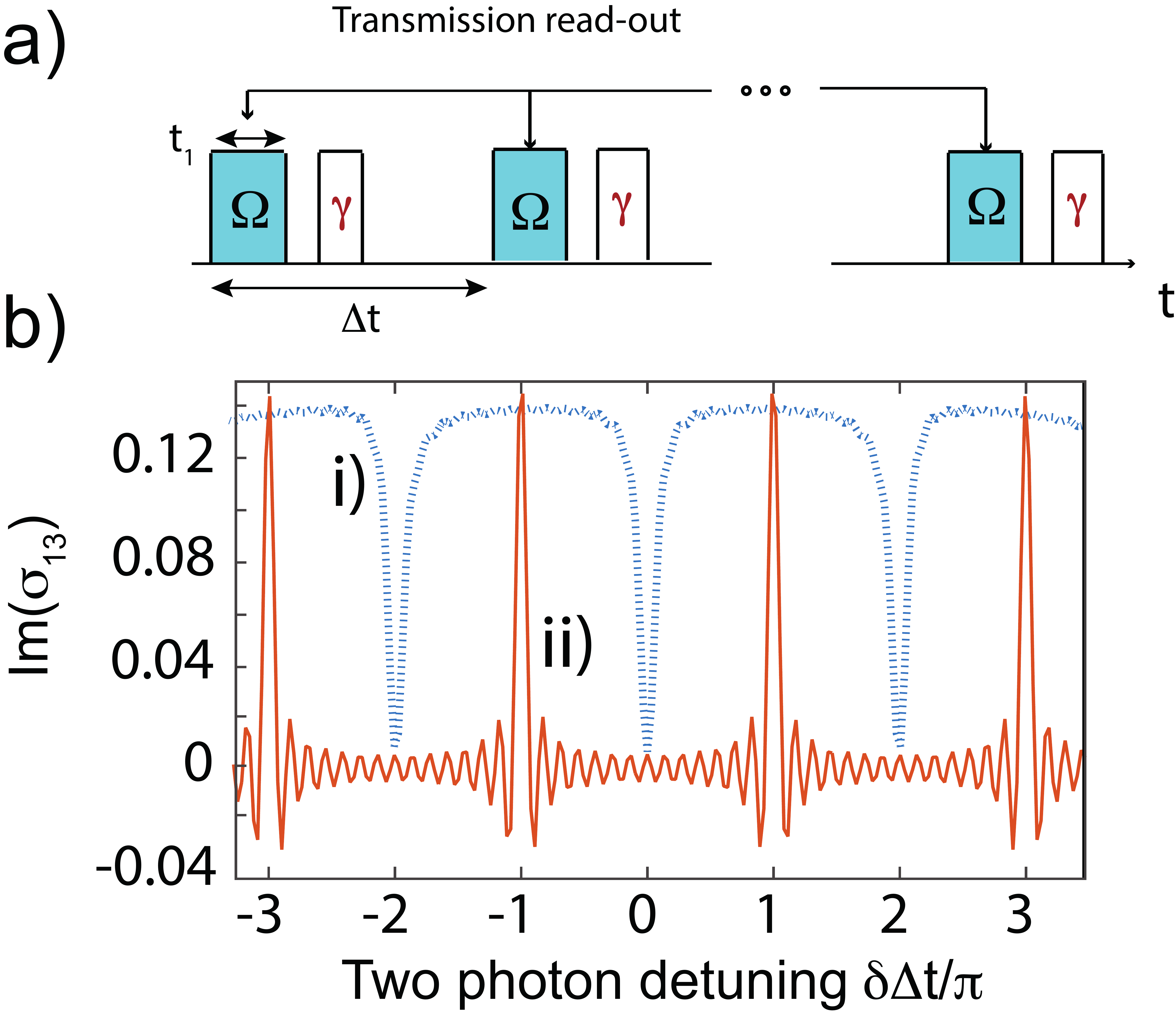}}}
\caption{
a) Pulse sequence used for studying the regime where the pulse area is $2\pi$.
b) Transmission spectra in the two different regimes: i) small area and ii) area close to $2\pi$ for the control pulse. The parameters are $\Omega_1/2\pi=0.03 $ MHz, $\Omega_2/2\pi=6.37$ MHz, $\Delta t=1\ \mu s$, $N=15$. Curve i) has been multiply by 1500 for clarity. 
}
\label{fig4}
\end{figure}

\section{Prospects for metrology} 

The narrow transmission lines in this stroboscopic state preparation has implications both in atomic clocks or magnetometry.
Coherent population trapping in fact already found use in precision measurement of atomic transitions \cite{Vanier2005}. 
\begin{figure}[h]
\centerline{\scalebox{0.35}{\includegraphics{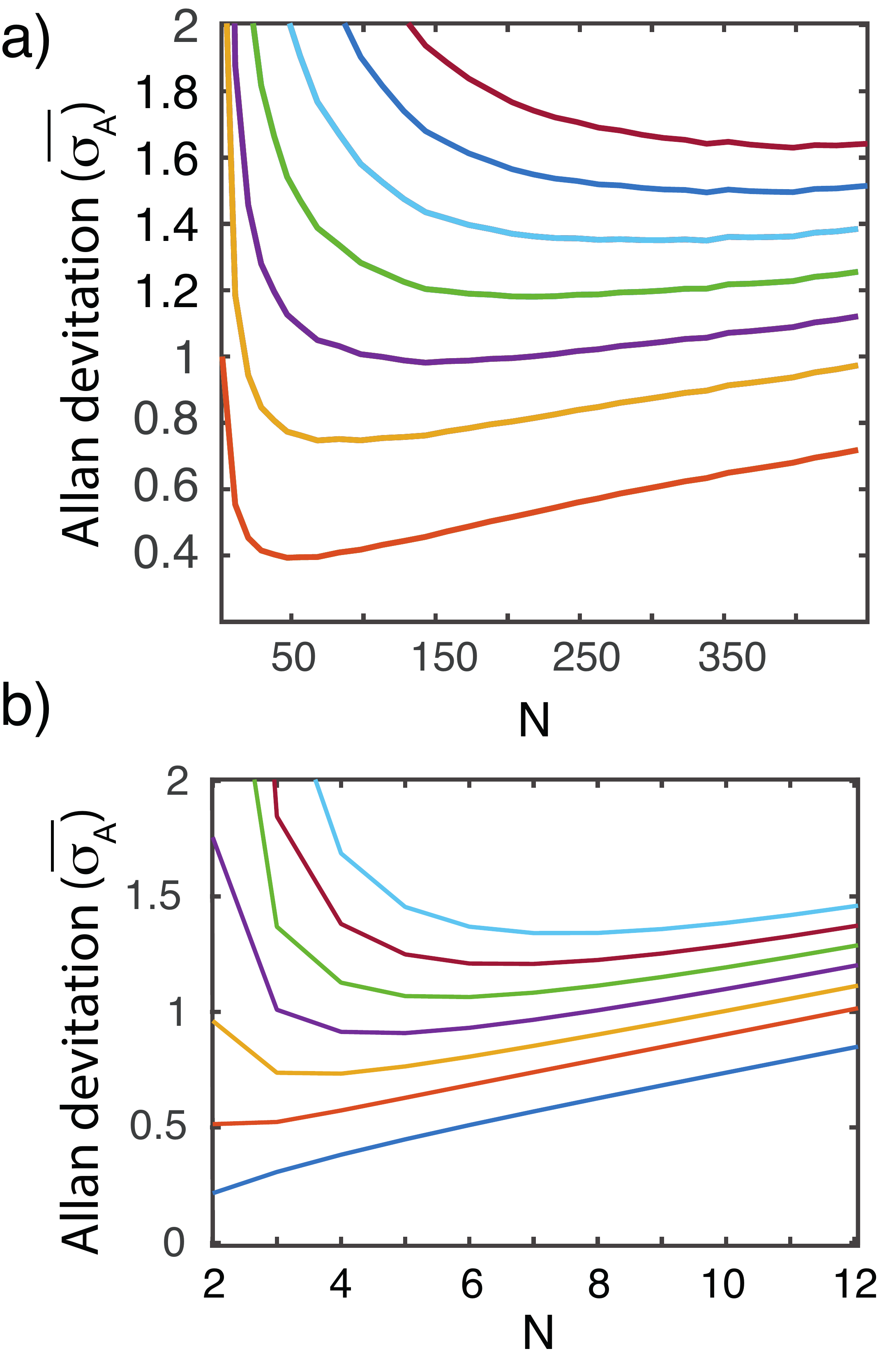}}}
\caption{a) Pulsed-CPT with controlled dissipation : Normalized Allan deviation as a function of the number of steps for $\gamma_0/2\pi$ ranging from 0 to 38.2 kHz in steps of 6.4 kHz (from bottom to top). $t_1=11$ ns. b) Ramsey-CPT : Normalized Allan deviation as a function of the number of steps for the same values of $\gamma_0/2\pi$. 
$\Omega_1/2\pi=\Omega_2/2\pi=6.37 $ MHz, $T=15\ \mu s$.
}\label{pCPTvsRCPT}
\end{figure}

Here we study the performance of the pulsed-CPT for metrology such as magnetic field or time measurements. Atomic clocks or magnetometers often use Ramsey sequences, where two $\pi/2$ pulses separated by a time $T$ are used to drive a two-level system.
In such a pulsed regime, scanning the laser frequency gives rise to a sinusoidal spectrum with periodicity $1/T$, providing a means to precisely estimate the atomic transition without power broadening. The fringe contrast will be given by the coherence time of the transition and so is the precision of the frequency measurement. 

Another protocol that minimizes power broadening together with the use of sub-natural lines is Ramsey-CPT \cite{Vanier2005}. The principle is to first prepare a dark state in the three-level system, to let the ground state coherence evolve "in the dark" and to measure the evolution of the ground state by applying a second pair of identical read-out pulses. If the lasers are two-photon detuned, the initial dark state population is transferred to the bright state so the fluorescence (or the absorption) is increased. This provides a means to read-out the atomic frequency close to the two-photon resonance condition. Here, due the multiple interference, the presented pulsed-CPT scheme will show even narrower lines so one could expect a high metrological performance. 

The frequency stability of clocks are typically characterized by the Allan deviation which, in the presence of white frequency noise, reads $$\sigma_A(\tau)=\frac{1}{\nu_{at}}\frac{\sigma_S}{\left(\frac{\partial S(\nu)}{\partial \nu}\right)_{\nu_m}}\sqrt{\frac{T_c}{\tau}},$$ 
where $T_c$ is the time of an interrogation cycle, $\tau$ is the averaging time, $\nu_{at}$ is the atomic resonance frequency, $\sigma_S$ is the noise of the measured signal $S$ and $\nu_m$ is the frequency at which the signal is measured. A small Allan deviation indicates a high precision in the frequency read-out. 
Magnetometers can be characterized in the very same way, the signal $S$ being proportional to the magnetic moment. 

Here, we compute $\sigma_A(\tau)$ for the pulsed-CPT and Ramsey-CPT schemes and quantify their performances as a function of the number of steps and for various decoherence rates in the ground state. To avoid optimization complications related to the choice of probe intensity, we chose to use fluorescence instead of transmission read-out for both schemes and use $\Omega_1=\Omega_2$ as in \cite{Jamonneau}.
Further, we use a pulse area far below $2\pi$ to be in the EIT regime. The analytical solutions presented earlier cannot be used here since we need to compute the excited state population instead of the coherence $|1\rangle\langle 3|$. We thus we resort to numerical simulations and use the XMDS package to solve the full Bloch equations \cite{DENNIS2013201}.

The fluorescence signal is acquired during each de-excitation pulse (except from the first one). The signal acquired during the $n^{th}$ decay process is directly proportional to the population at the end of the $n^{th}$ excitation pulse $\sigma_{33}(n\Delta t+t_1)$.
In order to make it independent on the actual experimental apparatus (collection efficiency, photodetector gain, number of atoms, ect.) the Allan deviation $\sigma_A$ is normalized to that of a two-pulse pulsed-CPT experiment without ground state decoherence (we write it $\sigma_A^{N=2}$).
For a fair comparison between the pulsed-CPT and Ramsey-CPT schemes, we use the same total sequence time $T_c$.
Increasing the number of steps within $T_c$ means that the signal will increase and that the precession time decreases. 
Note that more than 2 pulses may be required to optimize the Allan deviation in order to gather signal statistics. 

The populations are first prepared in the dark state.  
The Allan deviation (measured for the value of $\nu_m$ which minimizes it) is plotted in Figure \ref{pCPTvsRCPT}-a) as a function of the following number of steps for different $\gamma_0$. 
The results for $\overline{\sigma_A}=\sigma_A/\sigma_A^{N=2}$ are plotted for $\gamma_0/2\pi$ varying from $0$ to $38.2$ kHz in steps of 6.4 kHz. We observe that for each decoherence rate, an optimal number of steps $N_{\rm opt}$ is found that minimizes $\overline{\sigma_A}$.  This can be understood as follows : for $N < N_{\rm opt}$, the spectral lines are not optimally narrow because the multiple-interference is not at play. For $N = N_{\rm opt}$ however, we are in the multiple interference regime. 
These features are similar for all $\gamma_0$. As it can be expected also, the miminum of the Allan deviation decreases when $\gamma_0$ increases.
As the decoherence rate increases, more pulses are indeed needed to preserve the dark state so the optimal number of steps increases with $\gamma_0$.
When $N$ increases again above $N_{\rm opt}$, the precession time is reduced, therefore the linewidth increases and so does the Allan deviation (cf Fig.~\ref{pCPT}).

It is instructive to compare this scheme with the well-established Ramsey-CPT protocol.
Simulations for Ramsey-CPT are presented in Figure \ref{pCPTvsRCPT}-b), where the number of steps and decoherence rates in the ground state are also varied.  The main difference with pulsed-CPT with controlled dissipation is that the decay from the excited state to the ground state takes place at all times and that, for the Ramsey-CPT sequence, the photoluminescence signal is accumulated over all the excitation pulses \cite{Vanier2005}. We chose an area $\mathcal{A}$ which totally prepares the dark state at the end of each pulse. 
The results for the Allan deviation are shown for $\gamma_0/(2\pi)$ varying 
from 0 to 38.2 kHz in steps of 6.4 kHz (from bottom to top).
The trend for the Allan deviation is similar than for pulsed-CPT : when $\gamma_0$ increases, more pulses are needed to optimize $\sigma_A$ as a result of a compromise between signal strength and ground state coherence amplitude.
The optimum Allan deviation is on the same order of magnitude as for the pulsed-CPT with controlled decay, although it typically yields slightly smaller values.   

\section{Experimental implementation} 
The presented scheme was already realized in the microwave regime using NV centers in diamonds using coupled electronic and nuclear spins \cite{Jamonneau}.
It might be beneficial to use optical fields and enlarge the range of applications to, for instance, ultra-cold atoms. We now discuss a possible way to implement such a pulsed-CPT with controlled dissipation using neutral alkali atoms and trapped ions. 

\begin{figure}[ht]
\centerline{\scalebox{0.15}{\includegraphics{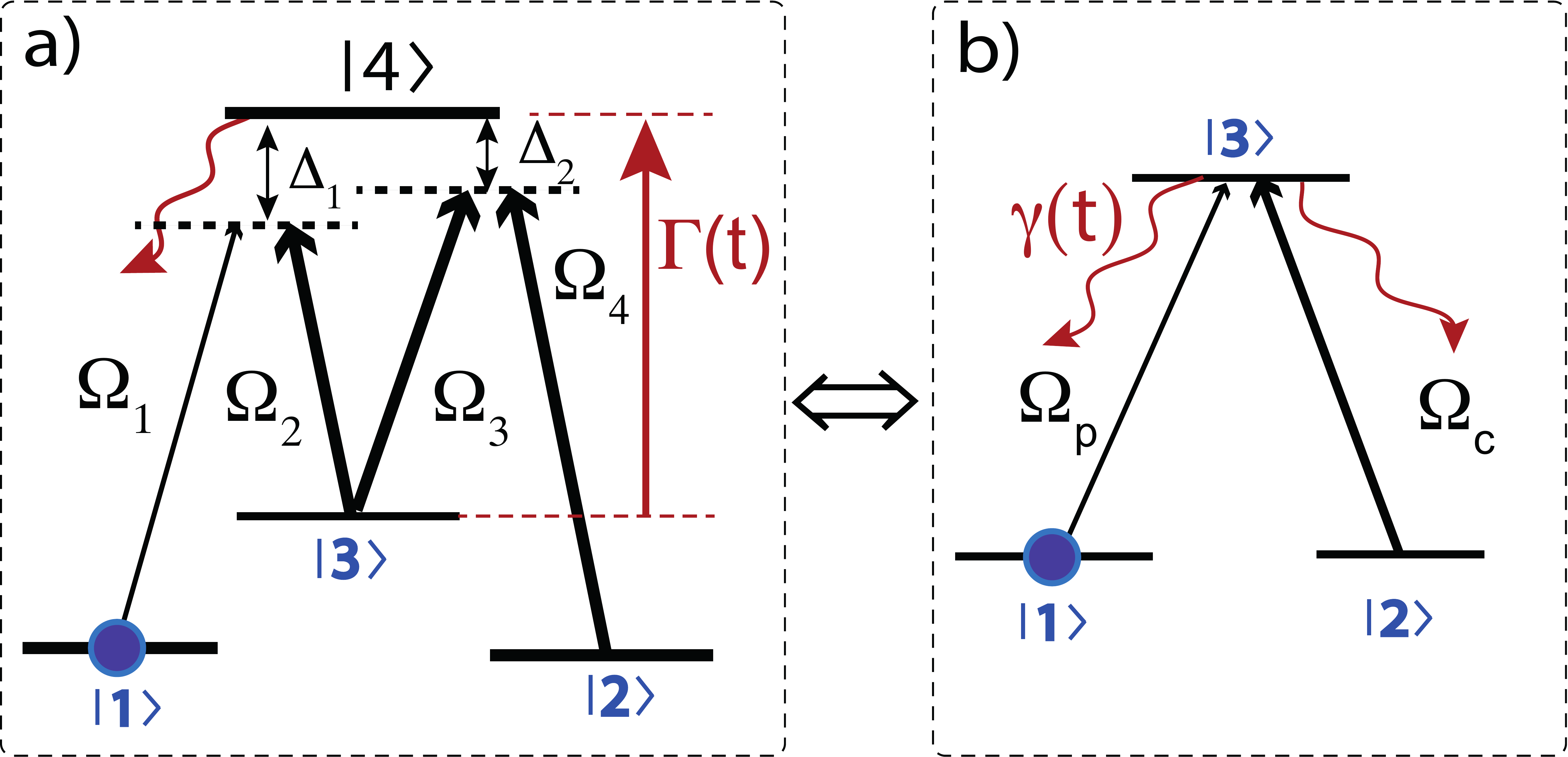}}}
\caption{a) Four-level scheme. b) The equivalent $\Lambda$ scheme.  When the four optical fields are off-resonantly coupled to  the level $\ket{4}$ and under two-photon resonance for the $\Lambda$ schemes $|1\rangle-|4\rangle-|3\rangle$ and $|3\rangle-|4\rangle-|2\rangle$, the scheme on the left is equivalent to the $\Lambda$ scheme shown on the right with effective Rabi frequencies $\Omega_{p,c}$. Controlled dissipation is done via Raman scattering with a pulsed laser with coupling $\Gamma$ tuned to the $\ket{3}$ to $\ket{4}$ transition (red arrow).}\label{fig8}
\end{figure}

{\it Alkalies :}
In order to realize this scheme with alkali atoms, one possibility is to engineer an effective three level system using four atomic levels and two Raman transitions as depicted in Fig. 6-a).
This can for instance be realized with all isotopes of Rubidium or Cesium on the $F\rightarrow F-1$ transitions of the $D_1$ and $D_2$ lines. Provided the four lasers are detuned by more than several excited state linewidths, we can adiabatically eliminate the state $|4\rangle$ and write $|\frac{\partial\sigma_{i4}}{\partial t}|\ll |\Delta\sigma_{i4}|$, where $i=1,2,3$ for all the optical transitions.  This means that two Raman transitions 
$|1\rangle-|4\rangle-|3\rangle$ and $|3\rangle-|4\rangle-|2\rangle$ coherently drive the long-lived transitions $|1\rangle-|3\rangle$ and $|2\rangle-|3\rangle$. 
The resulting effective Rabi frequencies on transition $|1\rangle-|3\rangle$ and $|2\rangle-|3\rangle$ are given by 
$\Omega_p=\Omega_1\Omega_2/\Delta_1$ and $\Omega_c=\Omega_3\Omega_4/\Delta_2$ respectively.
We also require the difference between the detunings $\Delta_1$ and $\Delta_2$ to be greater than the width of these two Raman transitions. This means that $\Delta_1-\Delta_2\gg(\Omega_p,\Omega_c$). If this condition is not satisfied, ground state coherence between the ground state $|1\rangle$ and $|2\rangle$ generated solely by fields 1 and 4 would take place.

This four-level scheme is then equivalent to the three-level system shown Fig. 6-b). The lifetime of the state decay from the state $\ket{3}$ can effectively be shortened via a spontaneous Raman scattering using a laser driving the transition between $\ket{3}$ and $\ket{4}$. The detection can be done by measuring the Raman scattered light intensity, or the laser transmission in an EIT-like experiment. Combined with this pulsed-CPT technique, efficient clocks or magnetometers can thus be realized. 

The coherence time in the two ground states $|1\rangle$ and $|2\rangle$ can be several milliseconds with trapped alkali atoms. 
Typical pulsed-EIT experiments require pulses that are shorter than the excited state decay time require femto-second lasers \cite{Soares2, Sautenkov, Campbell}. In comparison, the proposed pulsed-EIT can easily be observed on micro-second timescales so that light pulses can be generated simply using acousto-optic modulators. 
This in fact bears similarity with the experiment done in a Rubidium cell in \cite{Pinel}, where cavity-like features were observed using multiple interference on long-lived spin waves in a Gradient Echo Memory \cite{Hetet}. Operating on microseconds time scale may have important practical consequences. For instance, it could be experimentally feasible to shape the relative phase between the laser pulses after each step in order to generate amplification of the probe \cite{KOCHAROVSKAYA1992175} or to observe the step-by-step growth of the collective ground state spin wave amplitude via EIT. \\

{\it Trapped ions :}
Another possible experimental implementation of the pulsed-CPT scheme is using trapped ions on a quadrupolar transition. For instance, using $^{40}$Ca$^+$, this transition is driven by a 729 nm laser tuned to the $S_{1/2}$ to $D_{5/2}$. A $\Lambda$-system can be realized using the Zeeman sublevels $S_{1/2}(m=1/2)$ and $S_{1/2}(m=-1/2)$ and a single excited Zeeman levels in the $D_{5/2}$ manifold \cite{SchmidtKaler}. 
A laser at 854 nm, tuned to the $D_{5/2}$ to $P_{3/2}$ transition can then induce spontaneous Raman scattering back to the two ground states of $S_{1/2}$. Repeating sequences of 729 nm + 854 nm excitations thus realizes the pulsed-CPT protocol with controlled dissipation using the long-lived $D_{5/2}$ excited state. \\

{\it Differential AC-Stark shifts :}
One well-known issue with optical metrology is the differential AC-stark shift induced by other nearby electronic levels. 
This effect comes from the two optical fields that drive the $\Lambda$ scheme, which can off-resonantly couple to these extra levels and generate a fictitious magnetic
field proportional to the degree of circular polarization. This induces systematic shifts that are equivalent to a two photon detuning. 
Estimating precisely differential light shifts in the case of pulsed-CPT is, in general, complicated since they depend crucially on the level structure. Pulsed-CPT protocols have however been shown to be less prone to such shifts \cite{Pati}. Besides, other schemes, such as hyper-Ramsey interferometry \cite{Yudin} or Ramsey-comb spectroscopy technique \cite{Morgenweg, Morgenweg2} have been proposed to mitigate parasitic effects of the light shifts. 
Using these proposals in combination with pulsed-CPT and controlled dissipation may enhance the metrological precision and will be left to further studies.

\section{Conclusion}

We studied analytically the dynamics of a three-level system driven by a pulsed train of coherent fields in the presence of a controlled decay from the excited state. 
Formulas for the width of the transmission window were found and Autler-Townes doublets were recovered for pulse areas that are multiple of $2\pi$. We compared the Allan deviation of the protocol to the Ramsey-CPT scheme with metrological applications in mind and demonstrate a similar performance. We also showed that it is possible to observe the step-by-step growth of the dark-state amplitude using several experimental platforms. 
In general, this work adds a new dimension to CPT and EIT. 
Using more evolved pulse protocols, it may be possible to create ultra-narrow lines and use it for more efficient atomic clocks, magnetometers or for new light storage protocols \cite{Saglamyurek}.

\acknowledgements
We acknowledge helpful discussions with Thomas Zanon-Vilette. This research has been partially funded by the French National Research Agency (ANR) through the project SMEQUI. RC acknowledges the financial support from Fondecyt Postdoctorado No. 3160154 and the International Cooperative Program ECOS-CONICYT 2016 grant number C16 E04
\appendix

\section{Analytical solution}
\label{ap1}

We list here the full Bloch equations that are used to describe the pulsed-CPT scheme with a controlled  dissipation $\gamma(t)$ from the excited state. They read : 

\begin{equation}
\label{eqBloch}
	\left\{
	    \begin{array}{l}
 	       \dot{\sigma}_{13}=-(\frac{\gamma}{2}+i\delta)\sigma_{13}+i\frac{\Omega_1}{2}(\sigma_{11}-\sigma_{33})+i\frac{\Omega_2}{2}\sigma_{12} \\
 	       \dot{\sigma}_{12}=-(\gamma_0+i\delta)\sigma_{12}+i\frac{\Omega_2}{2}\sigma_{13}-i\frac{\Omega_1}{2}\sigma_{23}^* \\
 	       \dot{\sigma}_{23}=-\frac{\gamma}{2}\sigma_{23}+i\frac{\Omega_2}{2}(\sigma_{22}-\sigma_{33})+i\frac{\Omega_1}{2}\sigma_{13}^* \\
 	       \\
 	       \dot{\sigma}_{11}=\gamma\sigma_{33}+i\frac{\Omega_1}{2}\sigma_{13}-i\frac{\Omega_1}{2}\sigma_{13}^* \\
 	       \dot{\sigma}_{22}=\gamma\sigma_{33}+i\frac{\Omega_2}{2}\sigma_{23}-i\frac{\Omega_2}{2}\sigma_{23}^* \\
			 \dot{\sigma}_{33}=- \dot{\sigma}_{11}- \dot{\sigma}_{22} .
 	   \end{array}
	\right.
\end{equation}

\begin{figure}[h]
\centerline{\scalebox{0.25}{\includegraphics{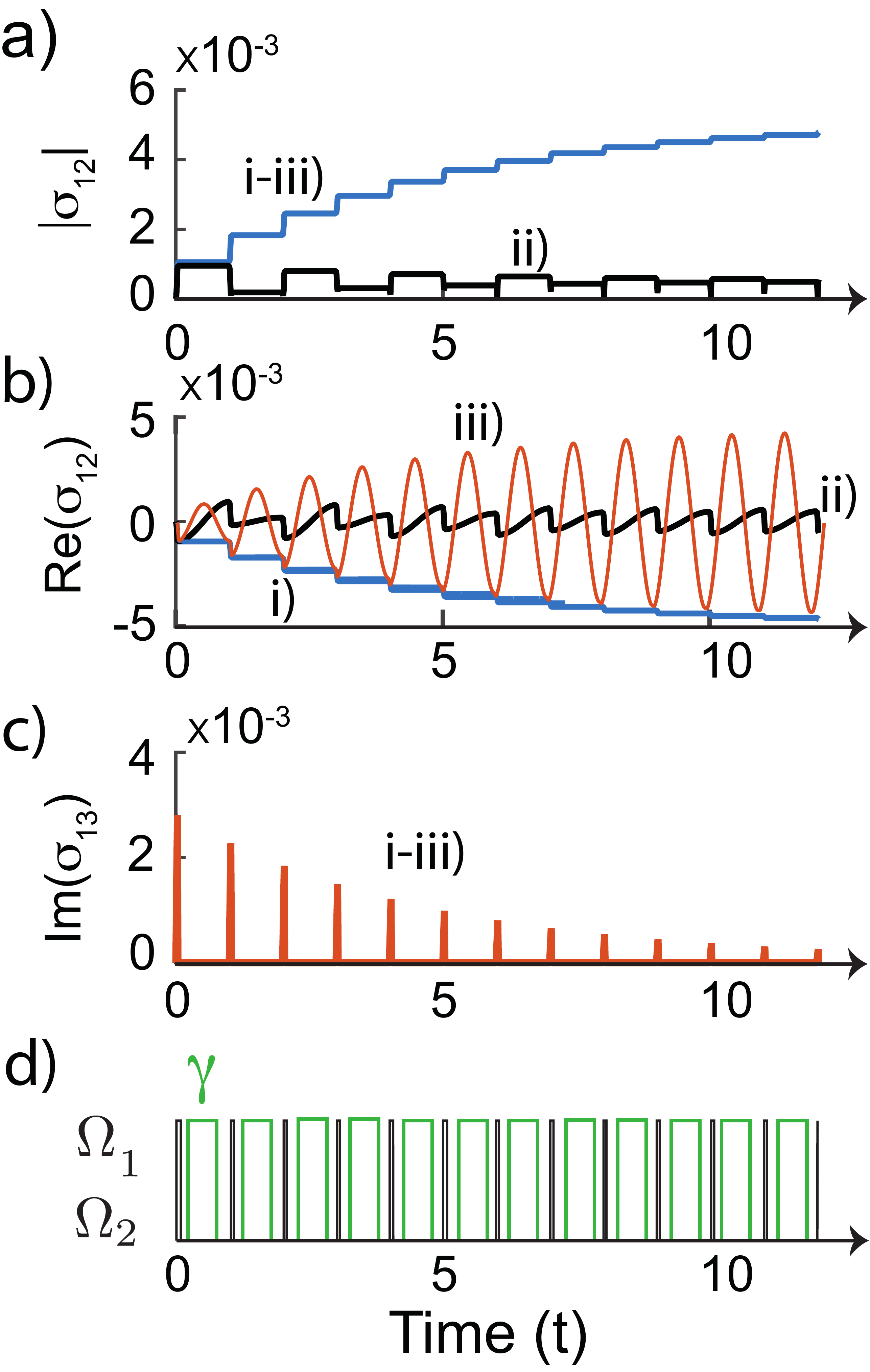}}}
\caption{a)-b) and c) Evolution of $\left|\sigma_{12}\right|$, Re$(\sigma_{12})$ and  Im$(\sigma_{13})$ for the detunings: i), $\delta=0$ ii), $\delta\Delta t=2\pi$ and iii) $\delta\Delta t=\pi$. Parameters: $\Omega_1/2\pi=31.8$ kHz, $\Omega_2/2\pi=6.37$ MHz, $\Delta t=1\ \mu s$ and $t_1=31$ ns ($\mathscr{A}=2\pi/5$). d) Excitation sequence. Note that i) and iii) are overlapping in Figure a) and c).}\label{fig44}
\end{figure}

In the following, we set $\gamma_0=0$. If the Rabi frequency of the field 1 is much smaller than the Rabi frequency of the field 2,
we have $\sigma_{11}\approx 1$, $\sigma_{22}\approx 0$ and $\sigma_{23}\approx 0$, so that 
\begin{equation}
\label{eqBlochsimp2}
	\left\{
	    \begin{array}{l}
 	       \dot{\sigma}_{13}=i\delta\sigma_{13}+i\frac{\Omega_1}{2}+i\frac{\Omega_2}{2}\sigma_{12}-\gamma\sigma_{13}  \\
 	       \dot{\sigma}_{12}=i\delta\sigma_{12}+i\frac{\Omega_2}{2}\sigma_{13} .\\ 
 	   \end{array}
	\right.
\end{equation}

The solution for the $n^{\rm th}$ step of the sequence is given by the equations 

\begin{equation}
\label{sol}
\sigma_{12}(t)=	\left\{
	    \begin{array}{l}
 	      e^{i\delta t}(A_ne^{i\frac{\Omega_2}{2} t}+B_ne^{-i\frac{\Omega_2}{2} t})+\alpha   \\
 	      \mbox{ \qquad \qquad if } t \in \left[(n-1)\Delta t,(n-1)\Delta t+t_1 \right] \\
 	      \\
 	      e^{i\delta(t-((n-1)\Delta t+t_1))}\sigma_{12}((n-1)\Delta t+t_1)\\
 	        \mbox{ \qquad \qquad if } t \in \left] (n-1)\Delta t+t_1, n\Delta t \right], \\ 
  	   \end{array}
	\right. 
\end{equation}
and
\begin{equation}
\label{sol2}
\sigma_{13}(t)=	\left\{
	    \begin{array}{l}
 	      e^{i\delta t}(A_ne^{i\frac{\Omega_2}{2} t}-B_ne^{-i\frac{\Omega_2}{2} t})-2\frac{\delta}{\Omega_2}\alpha \\
 	        \mbox{  \qquad \qquad \qquad si } t \in \left[(n-1)\Delta t,(n-1)\Delta t+t_1 \right] \\
 	      \\
 	      e^{(i\delta-\gamma)(t-((n-1)\Delta t+t_1))}\sigma_{13}((n-1)\Delta t+t_1)  \\
 	      \mbox{ \qquad \qquad \qquad si } t \in \left] (n-1)\Delta t+t_1, n\Delta t \right]. \\ 
  	   \end{array}
	\right. 
\end{equation}
Where $\alpha=\frac{1}{2}\frac{\Omega_1\Omega_2}{\delta^2-(\frac{\Omega_2}{2})^2}$.
The initial conditions  $\sigma_{12}(0)=0$ and $\sigma_{13}(0)=0$ give $A_1=\alpha\frac{\delta-\frac{\Omega_2}{2}}{\Omega_2}$ and  $B_1=-A_1-\alpha$.
The continuity of $\sigma_{12}$ and the hypothesis $\sigma_{13}(n\Delta t)=0$ for all $n$ is valid for full de-excitation and give the following expression for $A_n$ and $B_n$ \footnote{The recursive $A_n$ is of the form $u_{n+1}=cu_n+kd^n$}
\begin{equation}
\label{solAn}
	    \begin{array}{l}
			A_{n+1}=c^n\left( A_1+k\frac{\frac{d}{c}-\left(\frac{d}{c}\right)^{n+1}}{1-\frac{d}{c}}\right) 	 \\    
			\\
			B_{n+1}=   A_{n+1}e^{i\Omega_2 n\Delta t}-2\alpha\frac{\delta}{\Omega_2}e^{i(\frac{\Omega_2}{2}-\delta)n\Delta t} . \\  
	    \end{array}
\end{equation}

In the equation \ref{solAn}, $c=e^{-i\frac{\Omega_2}{2}\Delta t}\cos{(\frac{\Omega_2}{2} t_1)}$, \[k=\frac{\alpha}{2}\left(e^{i\delta(\Delta t-t_1)}-1+2\frac{\delta}{\Omega_2}(1-e^{-i\frac{\Omega_2}{2} t_1}e^{i\delta\Delta t})\right)\] and $d=e^{-i(\frac{\Omega_2}{2}+\delta)\Delta t}$.

The time evolution of the coherences is plotted in Fig.~\ref{fig44}-a), b), c) for a small pulse area and for three different detunings of $\Omega_1$. Trace i) corresponds to the resonance case $\delta=0$, trace ii), $\delta=\pi/\Delta t$ and trace iii) $\delta=2\pi/\Delta t$. For $\delta=0$ and $\delta=2\pi/\Delta t$, we observe that the coherences $\sigma_{13}$ tend to 0 when $N$ increases, which corresponds to the preparation of the dark state: after enough pulses $\Omega_1$ does not interact anymore with the atoms. We also see that for those two $\delta$, the atomic coherence reaches a steady state. 
We note that the evolutions of $\left|\sigma_{12}\right|$ are exactly the same for the two first detunings because they have the same phase and amplitude at the beginning of each pulse.

\begin{figure}[h]
\centerline{{\includegraphics{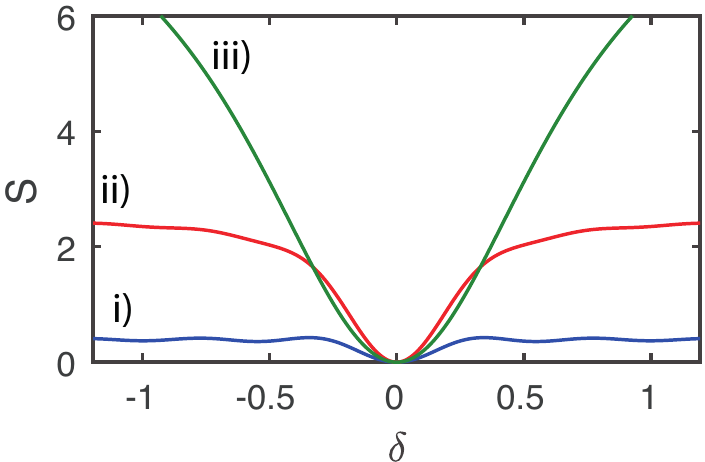}}}
\caption{Spectra $S$ for i) : $N=10<N_{opt}$ i). ii) $N_{opt}=54$. iii) $N=200>N_{opt}$. Parameters: $\Omega_1/2\pi=\Omega_2/2\pi=6.37$ MHz, $T=15\ \mu s$, $t_1=11 ns$ and $\mathscr{A}=0.2\pi$. }\label{pCPT}
\end{figure}

Three fluorescence spectra used to estimate the Allan deviation for the pulsed-CPT are plotted in Fig. \ref{pCPT}. They correspond to different sequences that have the same total length and pulse area but for different number of pulses. The signal is accumulated over all the de-excitation pulses except for the first one. Trace i), ii) and iii) show spectra for $N=10 $, $N=54$ and $N=200$ respectively.
For $ N < N_{\rm opt}$, 
the signal is not accumulated enough and so the photoluminescence rate drop is not steep yet because of the low number of excitation pulses.
For $N > N_{\rm opt}$, $\Delta t$ decreases as the number of pulses increases, which leads to a dilatation of the spectrum and thus reduces the slope of the spectrum close to $\nu_m$ (cf Fig. \ref{pCPT}). 

\end{document}